\documentclass[useAMS, usenatbib, usegraphicx]{mn2e}
\usepackage{rotating}

\def\d{\,d$^{-1}$}
\def\ge{$\gamma$\,Equ}
\def\deg{$^{\circ}$}

\title[CCD space photometry]
	{Reduction of time-resolved space-based CCD photometry developed for MOST Fabry Imaging data\thanks{MOST is a Canadian Space Agency mission, jointly operated by Dynacon Inc., the University of Toronto Institute of Aerospace Studies and the University of British Columbia, with the assistance of the University of Vienna, Austria.}
	}
\author[P.\,Reegen et al.]
	{P.\,Reegen$^{1}$\thanks{E-mail: reegen@astro.univie.ac.at}, 
	T.\,Kallinger$^{1}$,
	D.\,Frast$^{1}$, 
	M.\,Gruberbauer$^{1}$,
	D.\,Huber$^{1}$,
	\newauthor
	J.\,M.\,Matthews$^{2}$,
	D.\,Punz$^{1}$,
	S.\,Schraml$^{1}$,
	W.\,W.\,Weiss$^{1}$,
	R.\,Kuschnig$^{2}$,
	\newauthor
	A.\,F.\,J.\,Moffat$^{3}$,
	G.\,A.\,H.\,Walker$^{2}$,
	D.\,B.\,Guenther$^{4}$,
	S.\,M.\, Rucinski$^{5}$,
	\newauthor
	and D.\,Sasselov$^{6}$ \\
$^{1}$Dept. of Astronomy, University of Vienna, T\"urkenschanzstrasse 17, A-1180 Vienna, Austria \\
$^{2}$Dept. of Physics and Astronomy, University of British Columbia, 6224 Agricultural Road, Vancouver, BC V6T 1Z1, Canada \\
$^{3}$D\'epartement de Physique and Mont M\'egantic Observatory, Universit\'e de Montr\'eal, CP 6128, Succursale Centre-Ville, Montr\'eal, \\QC H3C 3J7, Canada\\
$^{4}$Dept. of Astronomy and Physics, St. Mary's University, Halifax, NS B3H 3C3, Canada \\
$^{5}$Dept. of Astronomy and Astrophysics, David Dunlap Observatory, University of Toronto, P.O. Box 360, Richmond Hill, \\ON L4C 4Y6, Canada \\
$^{6}$Harvard-Smithsonian Center for Astrophysics, 60 Garden Street, Cambridge, MA 02138
	}
\begin{document}

\date{Accepted  . Received  ; in original form 2005 January 1}

%\pagerange{\pageref{firstpage}--\pageref{lastpage}} \pubyear{2005}

\maketitle

%\label{firstpage}

\begin{abstract}
The MOST (Microvariability \& Oscillations of STars) satellite obtains ultraprecise photometry from space with high sampling rates and duty cycles. Astronomical photometry or imaging missions in low Earth orbits, like MOST, are especially sensitive to scattered light from Earthshine, and all these missions have a common need to extract target information from voluminous data cubes. They consist of upwards of hundreds of thousands of two-dimensional CCD frames (or sub-rasters) containing from hundreds to millions of pixels each, where the target information, superposed on background and instrumental effects, is contained only in a subset of pixels (Fabry Images, defocussed images, mini-spectra). We describe a novel reduction technique for such data cubes: resolving linear correlations of target and background pixel intensities. This stepwise multiple linear regression removes only those target variations which are also detected in the background. The advantage of regression analysis versus background subtraction is the appropriate scaling, taking into account that the amount of contamination may differ from pixel to pixel. The multivariate solution for all pairs of target/background pixels is minimally invasive of the raw photometry while being very effective in reducing contamination due to, e.g., stray light. The technique is tested and demonstrated with both simulated oscillation signals and real MOST photometry.

\end{abstract}

\begin{keywords}
methods: data analysis -- space vehicles: instruments -- techniques: photometric.
\end{keywords}

\section{Introduction}

	\subsection{Space photometry}
	
	The enormous potential of stellar photometry from space is finally being realised by the MOST (Microvariability \& Oscillations of STars) mission, a low-cost Canadian Space Agency (CSA) microsatellite which was launched in June 2003 (\citealt{Walker et al 2003}; \citealt{Matthews 2004}).

	This potential -- particularly for stellar seismology -- had been recognised for almost two decades (e.\,g., 
	%\citealt{Brown Cox 1986}, 
	\citealt{Mangeney Praderie 1984}; \citealt{Weiss 1993}), and many dedicated space missions have been proposed. Missions which reached a fairly advanced state of development but were not ultimately funded include: PRISMA \citep{Appourchaux et al 1993}, PPM \citep{Brown Torres Latham 1995}, STARS \citep{Fridlund et al 1995}, SPEX \citep{Schou et al 1998}, and MONS \citep{Kjeldsen et al 1999}. The status of the funding for the ESA mision EDDINGTON \citep{Favata Roxburgh Christensen-Dalsgaard 2000} is still not entirely clear.

	The first instrument designed and built for stellar seismology through photometry from space was EVRIS \citep{Vuillemin et al 1998}, a 10-cm telescope feeding a photomultiplier tube detector, mounted aboard the MARS-96 probe. Unfortunately, MARS-96 failed to achieve orbit. EVRIS was intended to be the precursor to COROT \citep{Baglin et al 2004}, the CNES-funded mission due for launch in 2006, to explore stellar structure through seismology and search for planets through photometric transits. Expected to join COROT in space in 2008 is KEPLER \citep{Borucki et al 2003}, whose primary goal is detection of Earth-sized planets via transits, but it will also obtain photometry uniquely powerful for stellar astrophysics.

	Another useful tool for space photometry has turned out to be NASA's WIRE satellite, whose primary scientific mission of infrared mapping failed. However, the satellite has proved to be a stable functioning platform for its 5-cm startracker telescope and CCD, which has been exploited successfully for stellar photometric studies (e.\,g., \citealt{Buzasi et al 2000}).

	MOST, WIRE, COROT, KEPLER and EDDINGTON are all CCD-based photometric experiments, and the first three are low-Earth-orbit (LEO) missions with similar orbital environments (radiation and scattered Earthshine). All these missions have a common need to extract information on stellar variability from data cubes consisting of upwards of hundreds of thousands of two-dimensional CCD frames (or sub-rasters) containing from hundreds to millions of pixels each. The modes of observation range from in-focus (KEPLER) and defocussed imaging (COROT, and MOST in its Direct Imaging mode) of fields with many targets to Fabry Imaging of the instrument pupil illuminated by a single target (MOST in its principal operation mode).

	In addition to its scientific value, MOST is a superb testbed for LEO space photometric techniques which can be applied to other missions. We present here a comprehensive approach for handling and reducing techniques which are relevant for other LEO space photometry missions and ground-based CCD photometry of cluster fields.

	\subsection{MOST mission overview}
	
	Because MOST is the first fully operational CCD space photometer, we use its archive to describe our reduction technique. This justifies a brief description of the mission.
	
	The MOST instrument is a 15-cm Rumak-Maksutov optical telescope feeding twin CCD detectors (one dedicated to stellar photometry, the other for guiding) through a single broadband filter. MOST was designed to obtain rapid photometry (at least one exposure per minute) of bright stars ($V < 6\fm5$) with long time coverage (up to 2 months) with high duty cycle. Its goal is to achieve photometric precision down to a few micromagnitudes ($\mu$mag) in a Fourier amplitude spectrum at frequencies down to about 1 mHz. It has achieved this goal for targets such as Procyon \citep{Matthews et al 2004} and $\eta$ Boo \citep{Guenther et al 2005}.
	
	MOST is a microsatellite with a mass of only 54 kg and hence little inertia. It is able to perform optical photometry of point sources thanks to a stabilisation system or Attitude Control System (ACS) developed by Dynacon, Inc. (\citealt{Groccott Zee Matthews 2003}; \citealt{Carroll Rucinski Zee 2004}). Previous microsats could not achieve pointing stability better than about $\pm 1$ to $2^{\circ}$, essentially useless for optical astronomical imaging. The requirement for the MOST ACS performace was $\pm 25\arcsec$; the goal was $\pm 10\arcsec$. The actual ACS performance was improved during the Commissioning and early science operations of MOST through software upgrades and refined use of the ACS reaction wheels, so that pointing precision is now about $\pm 1\arcsec$ rms.
	
	Anticipating image wander across the MOST Science CCD as large as $\pm 25\arcsec$, and the lack of an on-board flatfielding calibration system (due to power and cost limitations), the MOST design relies on producing a pupil image illuminated by target starlight. This extended image is produced by a Fabry microlens (similar in concept to the Fabry lens common to photoelectric photometers) and moves by no more than 0.1 pixel if the star beam moves by $25\arcsec$ from optimum pointing. (The resolution of the CCD is $3\arcsec$ per physical pixel.) To provide simultaneous sky background measurements, and for redundancy against Fabry lens and/or CCD defects, MOST is equipped with a $6\times 6$ array of microlenses, each of which images the telescope entrance pupil. Centred on each microlens is a field stop, 1 arcmin in diameter, etched in a chromium mask. We refer to \citet{Walker et al 2003}, for a detailed description of the focal plane arrangement and in particular to Figs.\,7 and 8 of that paper.
	
	\begin{figure}
		\includegraphics[height=240pt, angle=90]{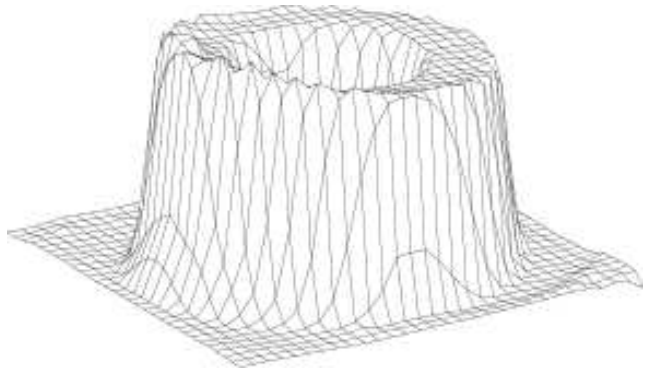}
		\caption{Mean normalised Fabry Image obtained from observations of \ge\ with Fabry lens 3/3.}
		\label{f:meanfabryimage}
	\end{figure} 
	
	The Primary Science Target is typically centred on one of these microlenses. A $2\times 2$ binned mean profile of the resulting Fabry Images is shown in Fig.\,\ref{f:meanfabryimage}. Each Fabry Image is an annulus with an outer diameter of 44 (unbinned) pixels. The pixels outside the annulus are sampled in a square subraster which gives an indicator for light scattered away from the pupil image or entering the focal plane independently of the telescope optics.
	
	The principal data format for MOST photometry -- referred to as Science Data Stream 2 (SDS2) --  consists of a resolved image of the Target Fabry Image, binned $2\times 2$ (to satisfy the limited MOST downlink rates), and 7 more heavily binned adjacent Sky Background Fabry Images. There are also subrasters of the CCD shielded from light which provide dark and bias measurements. Each SDS2 data file also contains a complete set of exposure parameters (exposure start time, integration time, CCD gain, etc.) and extensive satellite telemetry (e.\,g., CCD temperature, ACS errors, etc.).
	
	For longer-term data backup or to increase the sampling rate without exceeding the downlink limits, the MOST subraster images can be processed on-board, to produce Science Data Stream 1 (SDS1). Fabry Image Windows are integrated to give total intensities, backgrounds are generated from the four corners of each window, and sums of the pixel intensities binned in columns and rows are generated. SDS1 data can be sent to Earth at a rate 10 times higher than SDS2, but do not allow as flexible reduction and analysis.
	
	MOST also obtains Direct Imaging photometry of $1-6$ stars\footnote{Not recording Fabry window data makes it possible to allocate increased subraster area in the Direct Imaging Field.  In this case, the number of Direct Imaging Targets can be much larger than $6$, as in a cluster field like M\,67.}, based on defocussed images (FWHM $\sim$ 2.2 pixels) in portions of the Science CCD not covered by the Fabry microlens field stop mask. Processing of these data is described by \citet{Rowe et al 2005}. The ACS CCD is now also used to obtain photometry of about $20-30$ guide stars, to be described by \citet{Kuschnig et al 2005}.
	
	In this paper, we are primarily concerned with MOST photometry in the SDS2 format, although we do address SDS1 in Section \ref{sec:SDS1}.

\section{Error sources}						\label{sec:error}
%xxxxxxxxxxxxxxxxxxxxxx

	\subsection{Attitude Control System (ACS)}				\label{ssec:acs}
	
	The MOST Fabry Imaging approach is quite insensitive to telescope pointing errors, by keeping an extended nearly fixed pupil image on the same CCD pixels (to within 0.1 physical pixel if pointing deviations are less than $25\arcsec$). However, movements of the incoming beam of starlight on the Fabry microlens will introduce small photometric errors due to slightly different throughputs with changes in the light path through the telescope and camera optics. The greatest sensitivity may be due to tiny inhomogeneities in the Fabry microlens itself. The microlens array was etched into the BK-7 glass window above the CCD, and each element was tested in the lab for optical quality with artificial star images. Defects in certain microlenses in the array were identified and mapped in advance of launch, but there is always the possibility of subtle damage in orbit.
	
	One would expect some correlation between photometric errors and Attitude Control System (ACS) errors, but the relationship is complicated by the fact that the ACS system updates its guiding information typically once per second, whereas science exposures last for tens of seconds for the majority of targets. Thus, the star beam has moved across some complex path during the exposure. In an observing run, occasionally there is a pointing error which places the star too close to the edge of the field stop during part or all of an exposure, and significant signal from the stellar Point Spread Function (PSF) is lost.

	Pointing errors can also bring other sources (stars, galaxies and nebulae) near the Primary Science Target in and out of the field stop, introducing photometric errors. This was explored extensively in MOST mission planning, and there are no sources close to the MOST Fabry Imaging targets bright enough to affect the intended photometric precision. 

	In any event, MOST pointing has improved to the level where these factors are now relatively unimportant, but some of them did manifest themselves in the Commissioning Science data (e.g., $\kappa ^1$ Cet; see \citealt{Rucinski et al 2004}) and in some of the early Primary Science data.

	The reduction procedure takes ACS output into account to automatically reject images where at least one error value exceeds a pre-defined limit (see \ref{ssec:rejection}), indicating a target position being outside the nominal Fabry lens area at least during part of the integration.
	
	\subsection{Cosmic rays}						\label{ssec:cosmics}
	
	\begin{figure}
		\includegraphics[width=240pt]{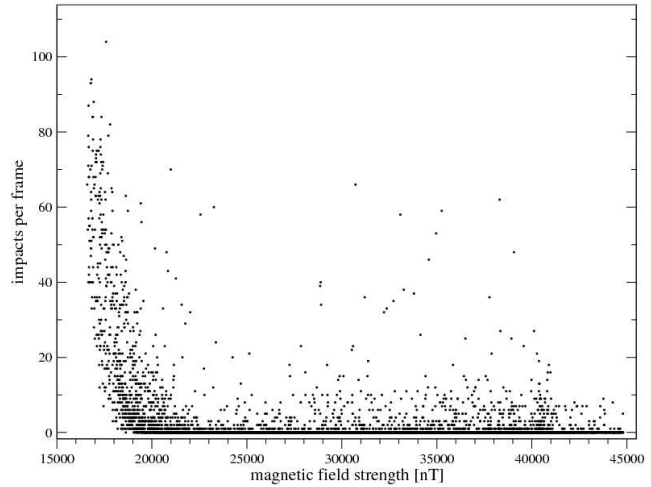}
		\caption{Impact density of cosmic rays vs. terrestrial magnetic field strength ($\kappa ^1$ Cet, October 2004). For transitions through the South Atlantic Anomaly, where the terrestrial magnetic field is weak, the impact rates rise significantly.}
		\label{f:saa}
	\end{figure} 
		
	Cosmic ray subtraction is important in ground-based CCD imaging, and it is generally even more critical for space-based photometry. The effects of energetic particle hits on the MOST Science CCD have been mitigated in the mission design in several ways:
	\begin{enumerate}
	\item The CCDs are shielded by the camera housing, equivalent to a spherical barrier of aluminium 5 mm thick.
	\item MOST is in a Low Earth Orbit (altitude 820 km) in which radiation fluxes are relatively low.
	\item MOST Fabry Imaging photometry employs only a small effective area of the Science CCD, rather than the entire $1024 \times 1024$ chip, presenting a relatively small target for incoming particles.
	\end{enumerate}

	MOST's orbit does carry it through the South Atlantic Anomaly (SAA, Fig.\,\ref{f:saa}), a dip in the Earth's magnetosphere which allows higher cosmic ray fluxes at lower altitudes. Passages through the SAA mean increased risk of an on-board computer crash due to a Single Event Upset (SEU) or ``latch-up''. MOST has experienced such particle-hit-induced crashes at an average rate of about once every two months in the 18 months of normal science operations. Otherwise, the increased particle fluxes during SAA passages do not have a significant effect on either ACS accuracy or stellar photometry.

	For those particles which do reach the relevant portions of the MOST Science CCD, there are two factors which make them easier to treat in the photometric reduction:
	\begin{enumerate}
	\item Cosmic ray hits are statistically independent in time (Poisson statistics, see \ref{sssec:coscand}).
	\item The most likely situation is that a particle will strike only a single pixel, making most hits easy to distinguish from other artefacts (see \ref{sssec:cosaffect}).
	\end{enumerate}

	\subsection{Stray light}					\label{ssec:stray light}
	
		\begin{figure}
			\vspace{5mm}
			\includegraphics[height=240pt, angle=90]{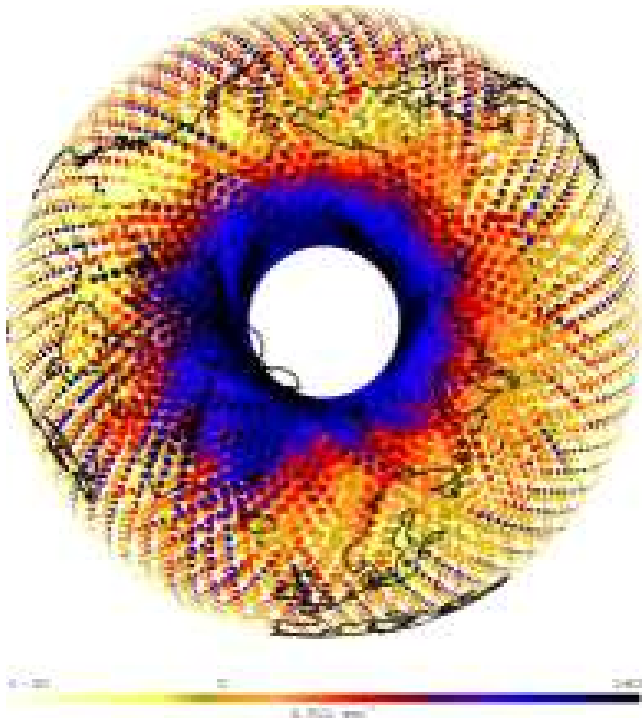}
			\caption{Residuals to a mean background intensity (invoked by terrestrial albedo variations) as a function of the sub--satellite point ($\gamma$ Equ data). The inner circle centred on the North Pole is void of information, since the maximum geographic latitude to be reached by MOST is $81.28^{\circ}$ \citep{Walker et al 2003}, not $90^{\circ}$. Spacecraft passages over Greenland are clearly associated with higher stray-light intensity, which introduces $1$\,d$^{-1}$ aliases to the Fourier spectra.}\label{f:stray light}
		\end{figure} 

	Any off-axis bright source such as the Earth will produce bright diffraction rims around the primary and secondary mirror even for perfectly baffled optics. These rims are clearly visible in Figs. \,\ref{f:rmsfabryimage} and \ref{f:rms_fourier_noise}. Since stray-light effects are imposed on us by nature and not peculiar to MOST, any follow-up space project will have to deal with stray-light contamination, and it is essential to find reliable correction techniques.
	
		\subsubsection{MOST orbit}		\label{sss:orbit}
	
		The MOST orbit plane inclination of 98.72\deg\ results in a dusk--dawn polar orbit because of the orbit plane precession due to the non-spherical terrestrial mass distribution. This precession compensates for the annual movement of our Sun along the celestial equator. The $\approx 9^{\circ}$ deviation of the orbit plane from orthogonality to the terrestrial equator together with the obliquity of the ecliptic of aproximately 23.5\deg\ results in passages of MOST over the illuminated north polar region during summer, which causes increased stray light. In extreme cases, the excessive stray light at certain MOST orbit phases can reduce the duty cycle from nearly $100\,\%$ to $60\,\%$.

		\subsubsection{Sunshine}
		
		For space imaging or photometry missions which need to point to fields at low solar elongations, scattered light from the Sun -- directly into the telescope or via the Zodiacal Light background -- would be the biggest challenge. The MOST mission was deliberately designed so that target fields are always in the opposite part of the sky from the Sun. The lowest solar elongation possible for MOST observations is about $135^{\circ}$. Direct solar light rejection is not a concern for MOST photometry.
	
		\subsubsection{Earthshine}

		For any LEO astronomical photometry satellite, scattered light from the illuminated portion of the Earth is a major source of sky background, and stray-light rejection is a major consideration in mission design and planning.

		Due to restrictions on the payload envelope in the MOST design, the instrument could not have a large external baffle. Three approaches were adopted to reduce the influence of scattered Earthshine on MOST  photometry:
		\begin{enumerate}
		\item The Sun-synchronous dawn-dusk orbit, combined with the choice of target fields in a Continuous Viewing Zone (CVZ) pointing away from the Sun, means that the MOST telescope tends to look out over the shadowed limb of the Earth.
		\item The telescope and camera are equipped with internal baffles and anti-reflective coatings.
		\item The spacecraft bus was intended to be a light-tight housing allowing light to enter only through the instrument aperture.
		\end{enumerate}
		None of these approaches is perfect.
		\begin{enumerate}
%		\item The MOST orbit plane has an inclination of 98.72$^{\circ}$, which is necessary to produce orbit  plane precession at the Sun-synchronous rate due to interaction with the J2 component of the Earth's rotational momentum. But because of the obliquity of the ecliptic is about 23.5$^{\circ}$, MOST orbits over the Earth's terminator only at times close to the Equinoxes. During the Solstices, a significant fraction of the bright Earth limb is visible to the MOST telescope.
		\item As was mentioned in Sec.\,\ref{sss:orbit}, the MOST orbit is Sun-synchronous. Since the obliquity of the ecliptic is about 23.5$^{\circ}$, MOST orbits over the Earth's terminator only at times close to the Equinoxes. During the Solstices, a significant fraction of the bright Earth limb is visible to the MOST telescope.
		\item The internal baffles and coatings appear to be working properly. But some light is scattered into the optics by parts of the the external door mechanism (which protected the MOST optics during launch and release into orbit and acts as a safeguard if there is a danger of the telescope pointing directly at the Sun), whose coatings may have been damaged on launch.
		\item One of the MOST spacecraft's external panels was ``shimmed'' when mounted to the bus (i.\,e., a thin piece of metal was inserted to compensate for a misalignment of mounting holes). This gap in the spacecraft structure allows light to enter the instrument from an unanticipated direction.
		\end{enumerate}

		As a result, the major component of sky background in MOST photometry is Earthshine, modulated with the satellite orbital period of 101.4 min (frequency = 14.2\,d$^{-1}$ = 165 $\mu$Hz) and, to a lesser extent, with a period of approximately 1 day as MOST's orbit returns it to a similar position over the Earth (and hence, a similar reflection feature in the terrestrial albedo) on a daily cycle. An example of the presence of stray light in raw MOST photometry is shown in Fig.\,\ref{fig:decorsteps}.

		The amplitude and shape of the stray-light modulation is highly dependent on the season of observations, the ``roll'' of the spacecraft (i.\,e., the position angle of the telescope about its optical axis), and the location of the target field within the CVZ. The relative contribution of maximum stray light also depends on the brightness of the target star. For Procyon \citep{Matthews et al 2004}, even though the observations are obtained near Winter Solstice and hence near the worst geometry for light scattered from the bright Earth limb, the stray light never exceeds $1\,\%$ of the stellar signal ($V \sim 0.3$). For the star $\gamma$ Equ ($V \sim 4.7$), observed near the Summer Solstice, the stray light reaches a maximum of about $1.75\times$ the stellar signal (see Fig.\,\ref{fig:decorsteps}).
		
		Fig.\,\ref{f:stray light} represents the variations of stray light as a function of MOST position over the Earth for many orbits during the $\gamma$ Equ run. The stray light reaches maxima over the North Pole and Arctic regions -- a combination of the high albedo of those areas which are continuously exposed to sunlight near Summer Solstice. (The Antarctic is in shadow, and does not dominate the stray light until times closer to the Winter Solstice.)
		
		In extreme cases, excessive stray light in MOST photometry can reduce its duty cycle from nearly $100\,\%$ to about $60\,\%$ per orbit. Such periodic gaps introduce aliases at a spacing of 14.2\,d$^{-1}$ or 165 $\mu$Hz. For comparison, the WIRE startracker (Buzasi et al. 2000) typically achieves an orbital duty cycle of about $20\,\%$, despite being in a similar orbit to MOST. This is because the WIRE platform must switch directions each half-orbit, to keep its batteries at the proper operating temperature, and the startracker focal plane receives even more severe scattered Earthshine. (Keep in mind that this instrument was designed as an ACS startracker, not a photometric instrument.)
		
		\subsubsection{Moonshine}

%\begin{table}
%  \begin{center}
%  \caption{Dates, angular separations, lunar phases, and illuminated fractions of the Moon disk for $\beta$ Vir observations (March, 2004).}
%  \label{t:betVirMoon}
%  \begin{scriptsize}
%  \begin{tabular}{lrrrr}
%  \hline
%  & \bf{JD$-$JD2000} & \bf{ang.\,sep.} & \bf{phase} & \bf{ill.\,frac.} \\
%  \hline  
%  Full Moon        & $1526.468056$ & $11\fdg 018$ & $0.000000$ & $1.000000$ \\
%  min.\,ang.\,sep. & $1527.233479$ &  $1\fdg 839$ & $0.025920$ & $0.993384$ \\ 
%  \hline
%  \end{tabular}
%  \end{scriptsize}
%  \end{center}
%\end{table}
 
		\begin{figure}
			\vspace{5mm}
			\includegraphics[bb=18 20 270 200, width=240pt]{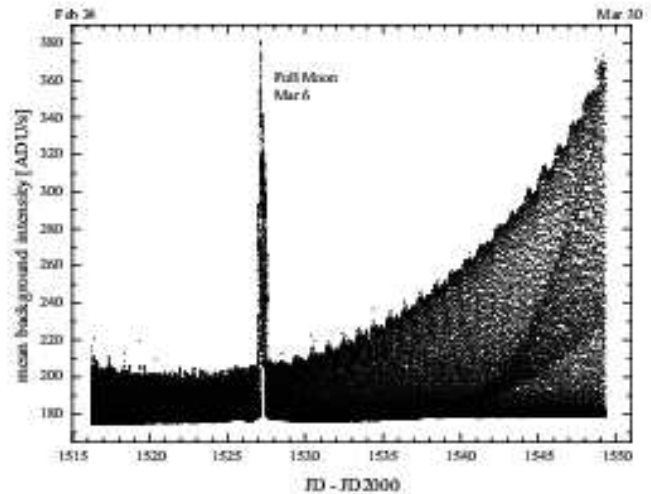}
			\caption{Mean background intensity vs. time for $\beta$\,Vir. Measurements on March 6, at a target position only $1\fdg 839$ from the Full Moon caused massive contamination by direct stray light.}
		\label{f:betVirMoon}
		\end{figure} 
	
		Because MOST targets are concentrated in a CVZ which is roughly equatorial ($36^{\circ} \leq \delta \leq -18^{\circ}$), the Moon can occasionally come very close to a MOST target field during observation. And because MOST observations are made in the anti-Sun direction, the phase of the Moon is close to Full when this happens.

		Two examples of a close passage of the Moon during MOST photometry occurred for the Commissioning Science target, ${\kappa}^1$ Cet, and the early Primary Science Target, $\beta$ Vir. In March 2004, the centre of the Moon came within $1.84^{\circ}$ of $\beta$ Vir during MOST monitoring of that star. The effects on the mean background are shown in Fig.\,\ref{f:betVirMoon}. The MOST stray-light rejection system was never designed to cope with such a close passage of such a bright source. Fortunately, lunar encounters like this are relatively rare and short-lived.

\begin{figure}
			\vspace{5mm}
			\includegraphics[bb=18 20 290 200, width=240pt]{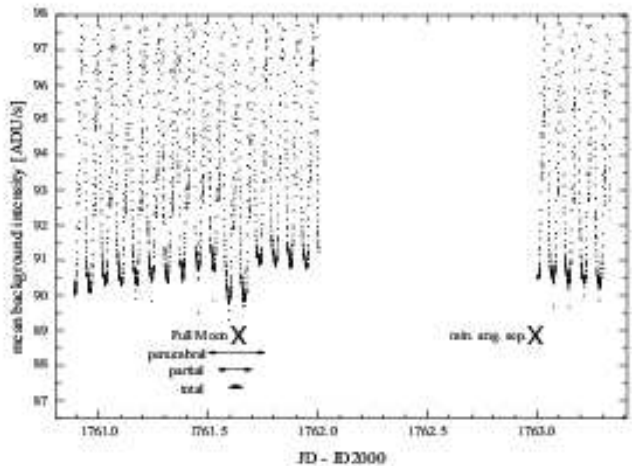}
			\caption{Mean background intensity vs. time for $\kappa ^1$ Cet. At an angular separation of 28 degrees, during an orbit phase of rather low Earthshine, lunar stray light begins to show up in the data. On Oct. 27, 2004, a lunar eclipse occurred during two orbits, producing a decay of stray-light intensity by $\approx 2\,\%$ at an angular separation of 20 degrees. The background intensity keeps increasing for two orbits after the eclipse, until the angle between Moon and target attains a minimum.}
		\label{f:lunareclipse}
		\end{figure}

		During observations of ${\kappa}^1$ Ceti in October 2004, the Moon came to within 20$^{\circ}$ of the star, and as it approached, it underwent a total lunar eclipse. This was an excellent opportunity to gauge the  contribution of Moonshine to MOST stray light at lunar elongations which would be somewhat more common in MOST observations. The mean background intensities for this encounter are shown in Fig.\,\ref{f:lunareclipse}.  

		This and other tests of MOST photometry with respect to lunar phase and elongation indicate that the Moon does not have a noticeable influence on MOST background measurements until it approaches within about  28$^{\circ}$ of the MOST target field. Generally, the Moon will affect the background at the level of a few percent. Although a typical MOST observing run is about a month long (covering only about one lunar synodic period), there is no evidence to date that the Moon introduces a periodic component to the MOST photometric background.
		
%\begin{table}
%  \begin{center}
%  \caption{Dates, angular separations, lunar phases, and illuminated fractions of the Moon disk for a lunar eclipse during $\kappa ^1$ Cet observations (October, 2004).}
%  \label{t:lunareclipse}
%  \begin{scriptsize}
%  \begin{tabular}{lrrrr}
%  \hline
%  & \bf{JD$-$JD2000} & \bf{ang.\,sep.} & \bf{phase} & \bf{ill.\,frac.} \\
%  \hline  
%  penumb.\,start  & $1761.503877$ & $20\fdg513$ & $0.995757$ & $0.999822$  \\
%  partial start    & $1761.551678$ & $21\fdg500$ & $0.997376$ & $0.999932$ \\
%  total start      & $1761.599630$ & $19\fdg519$ & $0.998999$ & $0.999990$ \\
%  max eclipse      & $1761.627847$ & $20\fdg850$ & $0.999955$ & $0.999999$ \\
%  Full Moon        & $1761.629167$ & $20\fdg798$ & $0.000000$ & $1.000000$ \\
%  total end        & $1761.656053$ & $18\fdg658$ & $0.000910$ & $0.999992$ \\
%  partial end      & $1761.703981$ & $18\fdg982$ & $0.002533$ & $0.999937$ \\
%  penumb.\,end    & $1761.751898$ & $19\fdg339$ & $0.004156$ & $0.999829$ \\ 
%  min.\,ang.\,sep. & $1763.008420$ & $15\fdg537$ & $0.046706$ & $0.978624$ \\ 
%  \hline
%  \end{tabular}
%  \end{scriptsize}
%  \end{center}
%\end{table}
 
	\subsection{CCD dark noise and temperature stability}
	
	The MOST CCDs are passively cooled to a temperature around $-35^{\circ}C$ and maintained at a given operating temperature by a trim heater with an accuracy of ${\pm}0.1^{\circ}C$.  Without regulation, the temperature would be modulated with the orbital period with a peak-to-peak amplitude of about $2^{\circ}C$. At temperatures below about $-30^{\circ}C$, the dark noise remains below $2$ electrons/pixel/sec and does not affect the Fabry Imaging photometry.

	The CCD focal plane temperature is monitored continuously to an accuracy of $0.01^{\circ}C$. All MOST photometry is checked for any correlation with CCD temperature and other operational parameters.

	\subsection{Fabry lens inhomogeneities}		\label{ssec:fabry}
	
	No system is optically perfect, and that is true for the Fabry microlenses as well.  If the star beam passes through a small flaw in the microlens, a distorted image of the entrance pupil will result. The deviation of an individual target image from the mean doughnut shape is used as a rejection criterion (see \ref{ssec:shape}). This also is an indicator of when the pointing has moved too close to the edge of the Fabry field stop and some of the starlight is obstructed.
		
\section{Concepts}							\label{sec:concepts}
%xxxxxxxxxxxxxxxxx

	\subsection{Nomenclature}					\label{ssec:annotation}
	
	Pixel intensities are considered to form a 3--dimensional array $I_{lmn}$, a data cube, where the index $n$ shall consistently refer to the time axis, and $l$ and $m$ represent the horizontal and vertical coordinates on the $n$th exposure. The number of exposures, i.\,e. the maximum value for the time index, is denoted $N$, the upper limits for spatial indices are $X$ (width) and $Y$ (height). In our case $X = Y$.
	
	In this context, indexing refers to virtual pixels, each virtual pixel intensity representing a $2\times 2$ bin of physical pixel intensities.
	
	\subsection{Statistical moments}		\label{ssec:moments}
	
	Statistical moments are denoted by an operator $\left<\:\right>$, where only indices inside the bracket are used for summation, and the order of the moment is given as an exponent. E.\,g., the temporal variance of intensity for the pixel $\left( l,m\right)$ is denoted $\left< I_n2\right> _{lm}$. Multivariate moments are written in the same sense, e.\,g. $\left< I_{jk}I\right> _{lmn}$ denotes the covariance of pixel $\left( l,m\right)$ with all other pixels $\left( j,k\right)$ on the $n$th image in the time series of exposures.
	
	A special case is the computation of moments through summation over all target pixels. In this case the abbreviation $I_T$ is used instead of using two indices. Correspondingly, $I_B$ denotes the intensity summed over all background pixels.

	\subsection{Data smoothing and detection of outliers}		\label{ssec:smoothing}

	The idea of identifying and correcting outliers in all our applications is based on data smoothing by a distance-weighted average (DWA), similar to the Whittaker filter \citep{Whittaker 1923} or Hodrick-Prescott filter \citep{Hodrick Prescott 1997}.
	
	The DWA will consistently be applied if data are to be smoothed in the time direction, according to
	\begin{equation}\label{EQWhittakerTime}
	F_\nu\left( t\right) \equiv \frac{\sum_{n=1,t_n\neq t}^{N}\left| t - t_n\right| ^{-\nu}I_{lmn}}{\sum_{n=1,t_n\neq t}^{N}\left| t - t_n\right| ^{-\nu}}\, ,
	\end{equation}
	as well as for image coordinates, where we write
	\begin{equation}\label{EQWhittakerSpace}
	F_{\lambda}\left( x,y\right) \equiv \frac{{\sum\sum}_{\left( l,m\right) =\left( 1,1\right) ,\left( l,m\right)\neq\left( x,y\right)}^{\left( X,Y\right)}R_{lm}I_{lmn}}{{\sum\sum}_{\left( l,m\right) =\left( 1,1\right) ,\left( l,m\right)\neq\left( x,y\right)}^{\left( X,Y\right)}R_{lm}}\, ,
	\end{equation}
	with
	\begin{equation}
	R_{lm} \equiv \left[\left( x - l\right) ^2 + \left( y - m\right) ^2\right] ^{-\frac{\lambda}{2}}\, .
	\end{equation}
	For these filters, we permit $\lambda$, $\nu$ to be positive semi-definite, real numbers, which help to adjust the degree of data smoothing. If these parameters are zero, the filter functions reduce to unweighted arithmetic means, respectively.
	
	In our application, the DWA is used in the time direction to identify and correct pixel intensities contaminated by cosmic ray impacts (\ref{sssec:coscand}, \ref{sssec:cosaffect}). The residual correlation between the integrated light curve and spacecraft position (\ref{sssec:satpos}) is performed using a DWA in terms of sub-satellite geographic latitude. The spatial filter (in terms of image coordinates) was used in context with the local dispersion model (\ref{ssec:LDM}), which did not prevail in the final reduction procedure and is mentioned here for completeness. 

\section{Implementation}							\label{sec:strategy}
%xxxxxxxxxxxxxxxx

	\subsection{Definition of target and background pixels}		\label{ssec:tbdef}

	The first step of the data reduction procedure is the assignment of each pixel to either target or background, hence defining an aperture to be used for photometry. This procedure is performed interactively.  
	
	\subsection{Rejection of images with extreme ACS error values} \label{ssec:rejection}

	The ACS provides a set of $xy$-errors for each image. These error values are defined on the interval $\left[ -25\farcs 8,\,25\farcs 8\right]$ in discrete steps of $0\farcs 2$. If the true ACS error exceeds this range, the maximum value is returned, indicating that the target was outside the nominal Fabry lens area -- at least during some fraction of the integration time. Consequently, readings with extreme ACS error values like these are automatically identified and rejected by the software.
		
	\subsection{Rejection of images with irregular geometry}			\label{ssec:shape}
	
	Both pointing problems and inhomogeneities of the Fabry lens may cause the image to be deformed. A three-step procedure has been chosen to correctly identify images with a deviating shape.
	\begin{enumerate}
	\item\label{ssec:shape:step1}
	All intensities are normalised according to
	\begin{equation}
	J_{lmn} \equiv \frac{I_{lmn}}{\left< I_T\right> _n}\, ,
	\end{equation}
	where it is sufficient to consider target pixels $\left( l,m\right)$ only.
	\item\label{ssec:shape:step2}
	The mean normalised Fabry Image, $\left< J_n\right> _T$, is computed. Fig.\,\ref{f:meanfabryimage} displays the mean Fabry Image for $\gamma$\,Equ.
	\item\label{ssec:shape:step3}
	Then the condition to keep the $n$th image is 
	\begin{equation}
	\sqrt{\frac{1}{K_T}\sum_T\left( J_{nT}-\left< J_n\right> _T\right) ^2} < g\, ,
	\end{equation}
	$K_T$ denoting the number of target pixels. All exposures not satisfying this condition are rejected. The reduction procedure actually applied to the MOST observations uniquely uses $g \equiv 4$, consistent with a $4\,\sigma$ criterion. This choice is a result of pure practical experience.
	\end{enumerate}
	
\begin{figure}
		\includegraphics[width=240pt]{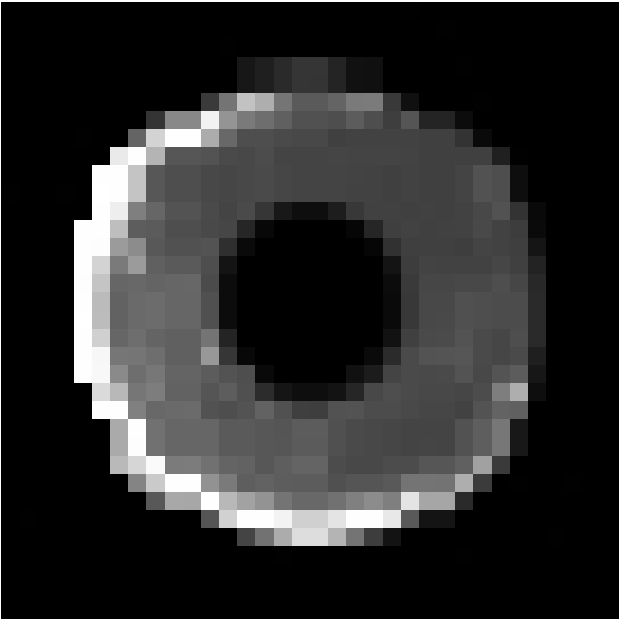}
		\caption{Rms deviations from the mean normalised Fabry Image in greyscale coding. These rms deviations are initial and decrease in the course of iterative rejections of poor-quality images.}
		\label{f:rmsfabryimage}
	\end{figure}

	Since the rejection of images in step \ref{ssec:shape:step3} influences the mean normalised Fabry Image, steps \ref{ssec:shape:step2} and \ref{ssec:shape:step3} are performed as a loop and terminated if no new rejection is performed in step \ref{ssec:shape:step3}. Fig.\,\ref{f:rmsfabryimage} represents the rms deviations, $\sqrt{\left< J_n^2\right> _T}$, for $\gamma$\,Equ in the first iteration. The inhomogeneous distribution of deviations on the subraster illustrates the importance of the pixel-resolved statistics described above.

	\subsection{Cosmic ray removal}						\label{ssec:cosmicscor}

		\subsubsection{Identification of candidate pixels} \label{sssec:coscand}
		
		Cosmic ray impacts are considered statistically independent and hence uncorrelated along the time axis. A cosmic ray hitting a given pixel does not influence subsequent readings of the same pixel. The principle of detecting cosmic-ray candidates is to compare each pixel intensity to one computed using a temporal DWA, which is conducted in three steps.
		\begin{enumerate}
		\item
		For a given pixel intensity $I_{lmn}$ at the time $t_n$, a smoothed intensity $F_\nu\left( t_n\right)$ is computed.
		\item
		The temporal rms deviation of the pixel intensities, $\left< I_n2\right> _{lm}$, is evaluated.
		\item
		The pixel $\left( l,m\right)$ is flagged as cosmic-ray candidate at the time $t_n$, if
		\begin{equation}
		I_{lmn} - F_\nu\left( t_n\right) > c\sqrt{\left< I_n2\right> _{lm}}\, .
		\end{equation}
		\end{enumerate}
		This procedure is conducted as a loop, only taking into account those readings of a pixel which have not been flagged as cosmic-ray candidates in the previous iterations. The loop terminates, if no new cosmic-ray candidates are found. Our application to MOST photometry consistently uses $c \equiv 4$. 
		
		\subsubsection{Correction of affected pixels}	\label{sssec:cosaffect}
		
\begin{figure}
			\includegraphics[width=240pt]{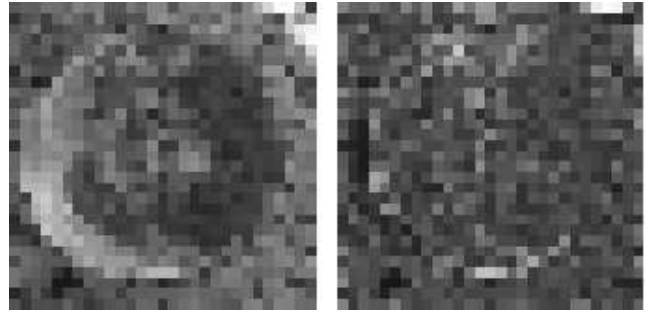}
			\caption{Detection rates for cosmic-ray candidates ({\em left}) and confirmed impacts ({\em right}) in greyscale coding ($\eta$ Boo, April 2005). The distribution of candidates reflects a misinterpretation of short-term stray-light effects, displaying a doughnut-shaped profile. As a significant improvement, the distribution of confirmed impacts is close to uniform, in agreement with the theoretical expectation for cosmic ray impacts.}
			\label{f:cosmics}
		\end{figure}

		Among the previously flagged cosmic-ray candidates, spatial information is used to decide whether a pixel represents a cosmic ray or not. The separation of cosmic rays and short-term stray-light variations is performed using the adjacent pixels on the same exposure. A cosmic-ray candidate is confirmed, if not more than $i$ adjacent pixels are cosmic-ray candidates, too. Fig.\,\ref{f:cosmics} displays the detection rates of cosmic-ray candidates and cosmic rays ($\eta$ Boo, April 2005). Whereas the candidate detection rates form a doughnut shape ({\em left}), which is due to the misidentification of short-term stray-light artefacts as cosmic-ray candidates, the rates of confirmed impacts ({\em right}) are nearly uniformly distributed over the fabry image. A value of $i = 1$ turned out useful for practical application. Only confirmed pixels are corrected.
		
		Intensities of pixels confirmed as cosmic rays are substituted by the value of the temporal DWA computed in \ref{sssec:coscand}.
		
		\subsubsection{Rejection of frames with high impact rates}	\label{sssec:cosrates}
		
		Evidently, each cosmic ray correction introduces additional errors to the data point referring to a considered frame. Hence, if the impact rate on a single frame exceeds a preselected threshold, the entire image is rejected. From our experience with MOST data, this threshold is set to 10.
		
	\subsection{Stray-light correction}          \label{ssec:strayl}

		\subsubsection{The decorrelation technique}	\label{sssec:decorrelation}
		
		The principle is to consider the stray-light sources as superposition of point sources. Each point source would produce a characteristic response function on the CCD image, and the combined stray-light contamination may be described as the integral of all these response functions. The resulting technique applied within our data reduction relies on consecutively resolving linear correlations between intensities of target and background pixels and will subsequently be called {\em decorrelation}.

		For a single point source, the shape of the related response function should be invariant to changing stray-light intensity, and the changes of pixel intensities are proportional to the stray light. The stellar signal will not influence the decorrelation procedure systematically unless its period and phase are too close to the orbit. In this case the period would be interpreted as stray light and removed. Depending on the brightness of the target as well as stray light and stellar amplitudes, this has massive implications on the detection of intrinsic stellar variability with a period comparable to the MOST orbit or a harmonic.

		\begin{figure}				
			\includegraphics[height=240pt, angle=90]{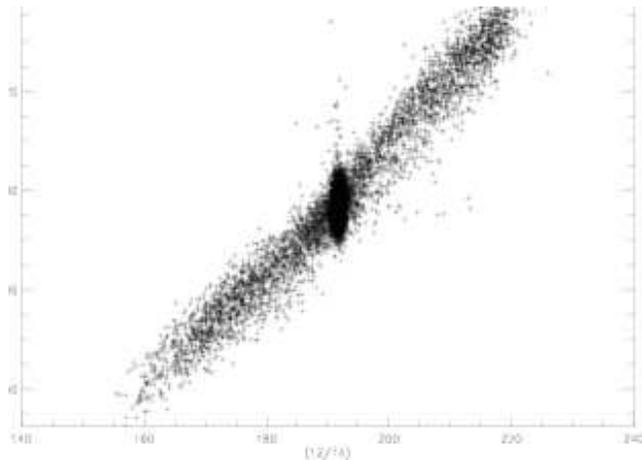}
			\caption{Correlation of the target pixel (20/20) with a background pixel (12/14) intensity for a 3-day subset of \ge\ observations.}\label{fig:corr}
		\end{figure}

		\begin{figure}%	bb=75 60 550 690		
			\includegraphics[height=240pt, angle=90]{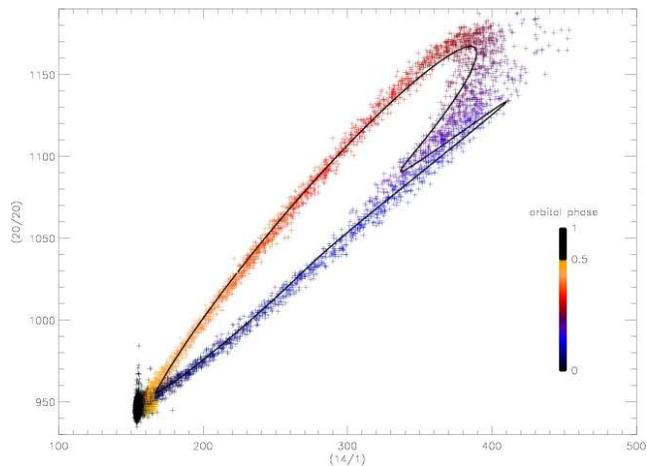}
			\caption{Same as for Fig.\,\ref{fig:corr}, but with background pixel (14/1). Colour coding refers to orbital phase, illustrating that the loop is associated with the motion of the spacecraft over one hemisphere and that (in this case) the satellite is susceptible to stray light depending on lateral illumination. The {\em solid line} refers to a modelled solution incorporating a stray-light pattern that moves across the detector, as described in \ref{sssec:lag}.}\label{fig:corrloop}
		\end{figure}
	
\begin{figure}				
			\includegraphics[width=240pt]{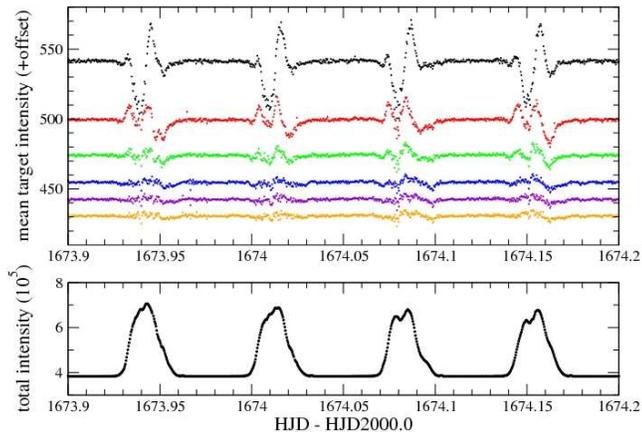}
			\caption{Light curve of \ge\ before stray-light decorrelation, computed as a sum of all 900 pixels ({\em bottom panel}). The decrease of stray-light contamination ({\em top panel}) after 1, 5, 10, 30, 60 and 95 decorrelation steps ({\em from bottom to top, with offsets applied for better visibility}) is shown for the mean of 421 target pixels.}\label{fig:decorsteps}
		\end{figure}

		Plotting the intensity of a target or background pixel versus that of a background pixel (where no stellar signal is present) yields a linear relation (Fig.\,\ref{fig:corr}). In case of one or more {\em extended} stray-light sources, the intensity-intensity diagram for two individual pixels may display a loop (Fig.\,\ref{fig:corrloop}), indicating an influence from several point sources, but dominating at different orbital phases, according to the possibilities of light being scattered through the aperture or lateral light leaks, and referring to the discussions of stray-light sources in \ref{ssec:stray light} and \ref{sssec:lag}. Hence, the contamination by stray light may be corrected by consecutively decorrelating a given pixel with respect to all background pixels. This is achieved by computing the slope of the linear regression in the intensity-intensity diagram. The correction of the linear trend is performed through preservation of the mean intensity. Fig.\,\ref{fig:decorsteps} illustrates how the background contamination due to stray light is being significantly reduced when correlating target pixels with (in the case of this figure, up to 95) background pixels.
	
		The portion of the stray light in the intensity-intensity diagram which cannot be removed entirely in one decorrelation step -- i.\,e. systematic deviations from the trend line, for example a loop -- is likely to re-appear as a linear trend when a different background pixel is used in a subsequent step. This property provides the chance to correct for most of the stray light without explicitly introducing the orbital period or phase. The present method turns out to be considerably less invasive towards the characteristics of spectral noise than a frequently used method based on the residual intensity determined from a phase plot with the orbital period.
	
		Finally, to take long-term stray-light changes into account, the complete dataset is divided into a sequence of subsets (e.\,g., three days long in the case of $\gamma$ Equ, Figs.\,\ref{fig:corr} and \ref{fig:corrloop}), and the stray-light correction is performed for these subsets individually.

		\subsubsection{Reduction of signal}
		
\begin{figure}				
			\includegraphics[width=240pt]{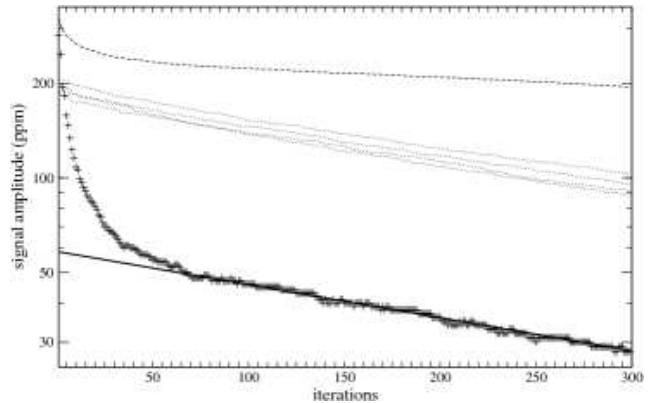}
			\caption{Decrease of amplitudes for the fundamental orbit period ({\em plus symbols}) and for four harmonic signals at 7, 23, 51 and 73\,d$^{-1}$ ({\em dotted lines}) artificially introduced into the Procyon data (Jan. to Feb. 2004), with increasing number of decorrelations. {\em Dashed line:} point-to-point scatter as a noise level estimator. {\em Solid line:} the linear decay of stray-light amplitudes at later decorrelation steps is exactly according to the trend in the amplitudes of synthetic signal. This property is used for reconstruction of the stellar signal.}\label{fig:lossosignal}
		\end{figure}

		In the optimum case, all of the stray light will be removed, but each decorrelation step necessarily reduces the signal in the target pixels. Signals of 200\,ppm amplitude at 4 different frequencies were additively synthesized in each of the target pixels. The decay of these amplitudes for the sum of all target pixels as a function of number of decorrelation steps is given in Fig.\,\ref{fig:lossosignal}. Obviously, the first decorrelation steps dramatically reduce the stray-light amplitude and after about 100 decorrelations there is a constant, but less significant decrease of the stray-light amplitude for each new decorrelation.

		But also the ``intrinsic'' signal is reduced in amplitude, independent of frequency. The amplitude decrease per decorrelation step is approximately constant for all decorrelation steps and corresponds to that for the stray light after many steps. This property allows one to reconstruct the initial amplitude even after a large number of decorrelations.
		
		This decrease in amplitude is due to the fact that -- in the case of pure noise -- the slope of a linear regression in the intensity-intensity diagram will scatter about zero, and that a trend correction will work for both positive and negative deviations. As a consequence, the decorrelation process with a non-zero slope always tends to reduce -- but never to increase -- the target pixel intensities. The expected value of the fraction of noise that is removed will be proportional to the rms deviation of the underlying probability distribution of regression slopes, where the terminology ``noise'' applies to signal at periods different from the orbital as well, if intensity is plotted vs. intensity.
		
		Simultaneously, the point-to-point scatter dramatically decreases with the first decorrelation steps, but shows asymptotically a shallower decay than the amplitudes. Hence, an optimum signal-to-noise ratio is obtained by terminating the decorrelation loop long before all iterations have been performed, and to use the later steps only to determine the slope of a linear trend for reconstructing the amplitude (solid line in Fig.\,\ref{fig:lossosignal}).
		
		The extent of the Fabry image is determined by the entrance pupil of the instrument, i.e. well-defined compared to direct-field images. However, we cannot be absolutely sure to have pure background pixels in what we consider as background. The question addresses a problem of aperture photometry in general, and possibly the decay of signal amplitudes with increasing number of iterations is -- to some extent -- due to target intensity leaking into our reference pixels (due to, e.g., scattered light in impurities of the microlens). The decorrelation technique is by far more sensitive to this leakage than ``classical'' subtraction of background intensity. A background pixel containing target information would cause the entire target signal to be removed in a single step of the decorrelation cascade. We would immediately see this effect, e.g. in Fig.\,\ref{fig:lossosignal}. The fact that the procedure works is the best indicator that the influence by the target on the background pixels is negligible.
		
		\subsubsection{Time lag}			\label{sssec:lag}

		A more subtle approach to the stray-light situation is to consider the projection of each point source moving across the detector, corresponding to the motion of the satellite above ground. This motion would cause two different pixels on the detector to be affected at different times, and the resulting time lag is expected to be larger for distant pixels.
		
		The formal consequence is to introduce a time lag between the correlation of two pixel intensities as a new free parameter and to perform an optimisation in the time direction through minimising the rms deviation of pixel intensities from the trend line in the intensity-intensity diagram (Fig.\,\ref{fig:corr}).
		
		We modelled theoretical intensity-intensity diagrams by computing the superposition of stray-light profiles of type
		\begin{equation}
		I \equiv A\,\mathrm{e}^{-\frac{1}{2}\left[\frac{\tan\left(\phi -\phi _0\right)}{\sigma}\right] ^2}\, ,
		\end{equation}
		$\phi$ denoting orbital phase, with adjustable amplitude $A$, orbital phase at maximum $\phi _0$, and width $\sigma$. A time lag is introduced by a different choice of $\phi _0$ for each of the two pixels compared.
		
		The solid line in Fig.\,\ref{fig:corrloop} represents a theoretical intensity-intensity diagram for a superposition of two stray-light patterns:
		\begin{enumerate}
		\item The loop is modelled by a stray-light bump sampled at both pixels with a lag of $1\fdg 44$ in terms of orbit phase. The width of the bump is set according to a FWHM in phase of $8\fdg 70$.
		\item A structure of FWHM $40^{\prime}$ in terms of orbit phase sampled at both pixels synchronously (i.\,e. without a time lag) produces the sharp peak below the loop. 
		\end{enumerate}
		This example illustrates the plausibility of moving stray-light patterns to be responsible for the loop-shaped intensity-intensity diagrams.
		
		However, when the idea was applied to MOST measurements, the increased number of free parameters did not lead to reasonable results. Although the postulate of static stray-light sources (permitting synchronous variations of pixel intensities only) is not claimed to be more reliable than the existence of moving stray-light artefacts -- since animated ``movies'' (using MDM, see Section \ref{sec:visual}) of the raw Fabry Images clearly show moving structures -- the quantitative description of this motion by a constant time lag between pixel intensities leads to a less efficient stray-light correction than the technique introduced in \ref{sssec:decorrelation} for mainly two reasons:
		\begin{enumerate}
		\item the number of decorrelation steps required to achieve optimum stray-light correction is not reduced substantially,
		\item the portion of signal removed with each decorrelation step (Fig.\,\ref{fig:lossosignal}) increases dramatically, and the final extrapolation to initial amplitudes leads to a higher noise level.
		\end{enumerate}

	\subsection{Choosing the optimum pixels in the Fabry Image} \label{ssec:pixsel}

	\begin{figure}
		\includegraphics[width=240pt]{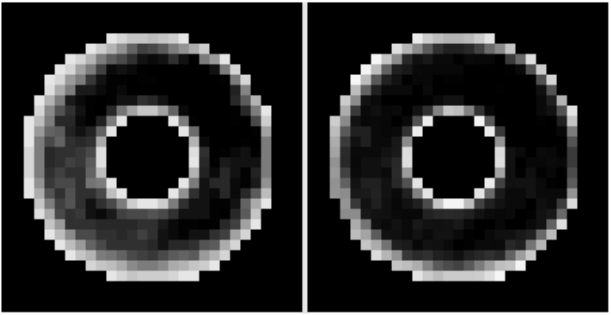}
		\caption{Noise maps of the stray-light corrected $\gamma$\,Equ data. 
		{\em Left:} relative rms error (rms error divided by the mean intensity of the pixel light curve). The higher relative scatter at the borders of the target image is due to photon noise, since the mean intensity in these regions is lower. In addition, some pixels inside the doughnut show a higher rms error than the rest.
		{\em Right:} amplitude noise level in the frequency range from 816 to 824\d. The absence of pixels with abnormally high spectral noise inside the doughnut indicates a stray-light reminiscence.}\label{f:rms_fourier_noise}
	\end{figure}

\begin{figure}
		\includegraphics[height=240pt, angle=270]{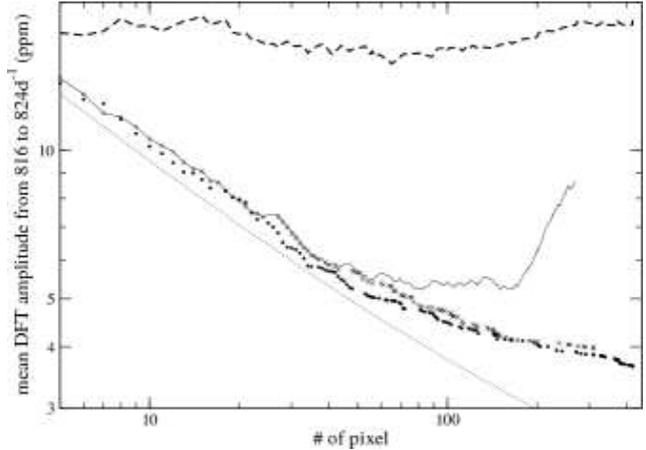}
	 	\caption{Dependence of the amplitude noise level (i.\,e. mean amplitude in a chosen frequency range) with increasing number of target pixels used for the $\gamma$\,Equ light curve. 
		{\em Solid line:} sorted by relative rms error. 
		{\em Crosses:} sorted by relative rms error and used only when improving the resulting amplitude noise. 
		{\em Bullets:} sorted by individual amplitude noise and used only when improving the resulting amplitude noise. 
		{\em Dotted line:} amplitude noise of the best target pixel divided by the square root of the number of used pixels -- a measure for the photon noise. 
		{\em Dashed line:} amplitude in the DFT spectrum at the orbit frequency divided by a factor of 20 for better visibility.}\label{f:compare_noise}
	\end{figure}

	As is illustrated in Fig.\,\ref{f:rms_fourier_noise}, the scatter of the target pixel light curves is distributed inhomogeneously in the Fabry Image. The left frame shows the distribution of the relative rms error (standard deviation divided by mean intensity) inside the doughnut. The annular areas with higher relative rms error are due to the lower mean intensity at the borders of the doughnut. The higher rms error in various areas inside the doughnut is due to remaining stray light, but also due to a higher point-to-point scatter in the pixel light curve. This is shown in the right frame of Fig.\,\ref{f:rms_fourier_noise} for the amplitude noise level of the target pixel light curves. The frequency range from 816 to 824\,d$^{-1}$ (the Nyquist frequency is $\approx 1400$\,\d\ for the $\gamma$\,Equ data) chosen for noise computation seems to be void of orbital, 1-day sidelobes, instrumental, and intrinsic peaks\footnote{Within these conditions, the actual choice of the frequency interval does not change the selection of target pixels substantially.}.
%	\begin{enumerate}
%	\item orbital,
%	\item 1-day sidelobe,
%	\item instrumental, and
%	\item intrinsic peaks\footnote{Within these conditions, the actual choice of the frequency interval does not change the selection of target pixels substantially.}.
%	\end{enumerate}
	As illustrated by Fig.\,\ref{f:compare_noise}, some target pixels are ``noisy'' enough to decrease the quality of the total light curve. Hence a proper selection of the target pixels used for the final light curve decreases the stray-light signal and the noise level in the frequency domain. 

	In the case of \ge, the solid line in Fig.\,\ref{f:compare_noise} indicates the evolution of the amplitude noise level (in the frequency range from 816 to 824\d) with increasing number of target pixels used for the resulting light curve . The target pixels are sorted by increasing relative rms value. 

	Obviously, the amplitude noise level in the amplitude spectrum decreases rapidly when adding the signal of the $\approx$\,50 ``best'' pixels. Adding more pixels of lower quality up to about 180 pixels does not improve the noise level in the resulting amplitude spectrum. Further increasing the number of pixels starts to increase the noise level in the resulting light curve. This is mainly due to pixels at the border of the doughnut with low signal and adverse photon statistics. 

	A closer inspection shows that occasionally the amplitude noise level increases although a rather good pixel has been added (e.\,g., close to the 20$^{\mathrm{th}}$ pixel). The reason for such a paradoxial result is random constructive interference leading to higher mean amplitude in the chosen frequency range, even if two low-noise datasets are combined. Using only those pixels which {\em improve} the resulting amplitude noise level leads to a significantly better quality of the final light curve (see crosses in Fig.\,\ref{f:compare_noise}).  
	
	The relative rms error -- as a superposition of signal and point-to-point scatter -- is obviously not the best criterion for the ``noise'' of a target pixel light curve. 

	Another examined sorting criterion is the mean amplitude level (here again in the frequency range of 816 to 824\d). The dots in Fig.\,\ref{f:compare_noise} display the evolution of this noise level when the target pixels are sorted by their individual amplitude noise and used for constructing the final light curve only if adding such a pixel improves the total light curve, i.\,e. reduces the amplitude level in the mentioned frequency interval. A total of 106 target pixel light curves were accepted in the given case, leading to an amplitude noise of $\approx 3.2$\,ppm in the relevant frequency range. However, there is no significant difference in the resulting light curve between amplitude noise and relative rms error in the time domain (crosses in Fig.\,\ref{f:compare_noise}) as sorting criteria. The resulting light curve is also insensitive to the choice of this frequency range, if void of stellar or instrumental signal. The dashed line in Fig.\,\ref{f:compare_noise} is associated the Fourier amplitude at the orbit frequency of $\approx 14.2$\,\d and illustrates that the influence of the choice of incorporated target pixels on the remaining stray-light peaks is negligible. In Fig.\,\ref{f:compare_noise}, this amplitude is rescaled by a factor $0.05$ for better visibility.

	\begin{figure}
		\includegraphics[height=240pt, angle=270]{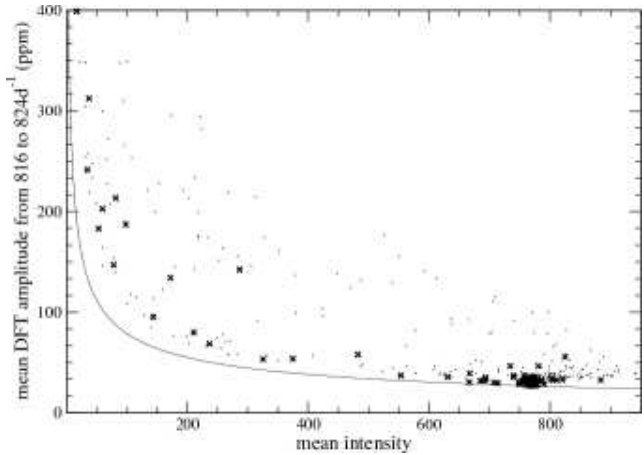}
		\caption{Amplitude noise level (i.\,e. mean amplitude in a chosen frequency range) versus mean intensity of the 421 target pixel light curves. ``$\times$'' symbols indicate pixels used for the final light curve. The line refers to the predicted photon noise for these 421 pixels.}\label{f:phot_noise}
	\end{figure}

	The dotted line in Fig.\,\ref{f:compare_noise} illustrates the theoretical improvement of the amplitude noise level with an increasing number of used target pixels assuming the same quality for all these pixels. Only the first (i.\,e. best) 30 target pixels appear to be photon noise limited. But what we see is caused by the different mean intensity level of the various target pixels. Fig.\,\ref{f:phot_noise} illustrates the dependence of the amplitude noise level of target pixel light curves on their mean intensity (dots). Those target pixels finally used are marked by ``$\times$'' symbols, and it is obvious that most (about $80\,\%$) of the target pixels are close to the photon noise limit.

	\begin{figure}
		\includegraphics[height=240pt, angle=270]{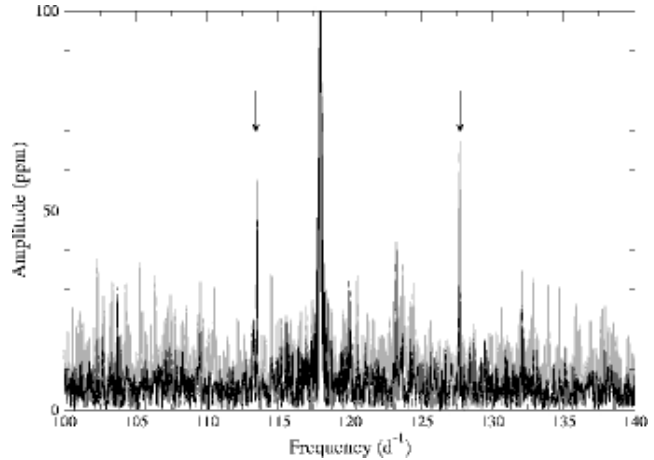}
		\caption{DFT spectrum of the final $\gamma$\,Equ light curve based on all ({\em grey}) and on the 106 selected target pixels ({\em black}). The arrows indicate the 7th and 8th harmonics of the orbit period.}\label{f:ampspec}
	\end{figure}

	The improvement of the light curve due to a proper pixel selection is illustrated in Fig.\,\ref{f:ampspec} by showing the amplitude spectrum of $\gamma$\,Equ in a frequency range with already reported pulsation modes. The grey line in the background indicates the DFT of the light curve based on all target pixels. This spectrum is crowded with instrisic, orbital overtone, and 1-day sidelobe peaks. The black line corresponds to the DFT of the light curve based on the selected target pixels and shows almost exclusively peaks not associated to any orbit harmonic. 

\subsection{Correlation with spacecraft parameters} \label{ssec:residcor} 

\begin{table}
   \caption{Comparison of stray-light amplitudes and frequency-domain noise between raw and reduced data for $13$ MOST targets observed from Oct 2003 to March 2005,
   {\em date:} time of observation (yymmdd);
   {\em target:} target name;
   {\em orbit (raw), orbit (red):} mean amplitude (mmag) of orbital frequency and harmonics up to order $9$, for raw and reduced data, respectively;
   {\em interval:} frequency interval (d$^{-1}$) used for calculation of amplitude noise;
   {\em noise (raw), noise (red):} amplitude noise (mmag), for raw and reduced data, respectively.}
   \label{TABquality}
  \begin{center}
  \begin{tiny}
  \begin{tabular}{@{}clrrcrr@{}}
  \hline
  \bf{date}&            \bf{target}&            \bf{orbit}&       \bf{orbit}&   \bf{interval}&
  \bf{noise}&     \bf{noise}\\
  &            	        &                       \bf{(raw)}&       \bf{(red)}&   &
  \bf{(raw)}&     \bf{(red)}\\
  \hline
  \\
  031008-031027&     $\delta$\,Cet&          $16.137$&                     $0.819$&                      $[158,170]$&
  $2.252$&                    $0.179$\\
  031028-031205&     $\kappa ^1$\,Cet&       $5.481$&              $0.588$&                      $[129,141]$&
  $0.255$&            $0.072$\\
  040103-040209&     Procyon&                $2.358$&              $0.054$&                      $[115,127]$&
  $0.820$&            $0.006$\\
  040223-040329&     $\beta$\,Vir&           $3.589$&              $0.075$&                      $[158,170]$&
  $0.088$&            $0.006$\\
  040329-040413&     $\tau$\,Boo&            $9.701$&              $0.075$&                      $[156,168]$&
  $0.213$&            $0.036$\\
  040413-040511&     $\eta$\,Boo&            $4.312$&              $0.110$&                      $[156,168]$&
  $0.062$&            $0.005$\\
  040517-040614&     $\zeta$\,Oph&           $42.475$ &            $0.662$&                      $[186,198]$&
  $0.130$&            $0.029$\\
  040728-040816&     $\gamma$\,Equ&          $32.352$ &            $0.282$&                      $[172,184]$&
  $0.153$&            $0.024$\\
  040830-040927&     51\,Peg&                $10.542$ &            $0.155$&                      $[187,199]$&
  $0.043$&            $0.007$\\
  041015-041104&     $\kappa ^1$\,Cet&       $2.189$ &             $0.155$&                      $[158,170]$&
  $0.021$&            $0.017$\\  
  041105-041204&     HR\,1217&               $22.163$ &            $0.448$&                      $[158,170]$&
  $0.211$&            $0.031$\\
  050124-050211&     Procyon&                $1.334$ &             $0.031$&                      $[168,180]$&
  $0.335$&            $0.004$\\
  050215-050322&     $\iota$\,Leo&           $1.806$ &             $0.053$&                      $[158,170]$&
  $0.106$&            $0.006$\\
%  050322 -- 050331&     $\beta$\,Com&           $3.68014$ &             $3.94599$&                      $[160,172]$&
%  $0.54904$&            $0.82336$\\
%  \\
%  050331 -- 050421&     $\eta$\,Boo&            $3.77445$ &             $4.83396$&                      $[158,170]$&
%  $0.35573$&            $0.56334$\\
%  \\
%  050421 -- 050515&     $\tau$\,Boo&            $33.78128$ &            $3.64837$&                      $[156.3,168.3]$&
%  $0.40764$&            $0.56274$\\
%  \\
  \hline
  \end{tabular}
  \end{tiny}
  \end{center}
\end{table}

	After correcting for stray light, which is the most important aspect of the data reduction for MOST, a weak correlation of the photometry with various spacecraft parameters may still be present. Among the variety of parameters available and examined, residual correlations with the photometry are only found for the sub-satellite geographic latitude, the CCD temperature (only before summer 2004), and the ACS deviation (only for Commissioning and very early Science targets).

%	\begin{enumerate}
%	\item the sub-satellite geographic latitude,
%	\item the CCD temperature (only before summer 2004), and
%	\item the ACS deviation (only for Commissioning and very early Science targets).
%	\end{enumerate}

		\subsubsection{Satellite position}	\label{sssec:satpos}

\begin{figure}
			\includegraphics[height=240pt, angle=270]{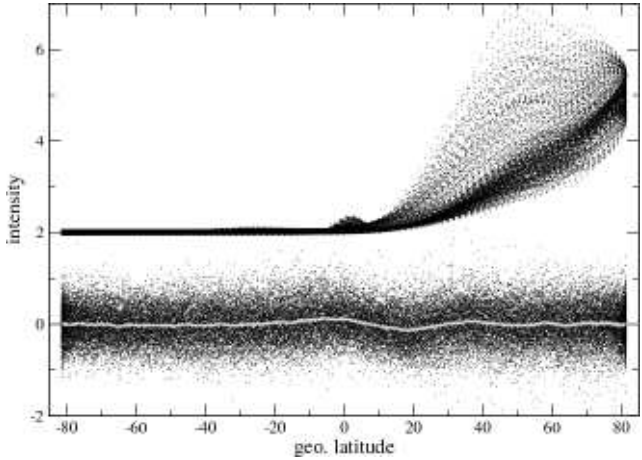}
			\caption{$\gamma$\,Equ photometry versus sub-satellite geographic latitude. \em Top: \rm mean target pixel intensities of the raw data (divided by 100 and offset by $2$\,ADU\,sec$^{-1}$ for better visibility) show a stong stray-light component at northern latitudes. \em Bottom: \rm stray-light corrected photometry after optimum target pixel selection. Only a marginal dependency of intensity on sub-satellite geographic latitudes remains (grey line).}\label{f:geoLat}
		\end{figure}

		In the case of $\gamma$\,Equ, stray-light correction and proper target pixel selection reduce the stray-light contribution to the photometric signal by a factor $\approx$ 1700, from $66\,\%$ to $0.04\,\%$. A comparison of the raw and corrected mean target pixel intensities is given in Fig.\,\ref{f:geoLat}. The top panel shows the raw data (mean of all pixels per frame, scaled to fit the figure size by dividing by 100, and offset by 2 ADU per second for better visibility) as a function of geographic latitude. The bottom panel displays the stray-light corrected photometry after stray-light correction and target pixel selection.

		Long-term variations were removed from the data by subtracting a moving average. Only a marginal dependence of intensity on the sub-satellite geographic latitude remains. This residual correlation is corrected by subtracting a DWA (grey line) in the intensity vs. latitude plot and has no influence on the amplitude noise level, but reduces the amplitudes of the higher orbit harmonics.

		\subsubsection{Timing effects}

\begin{figure}
			\includegraphics[height=240pt, angle=270]{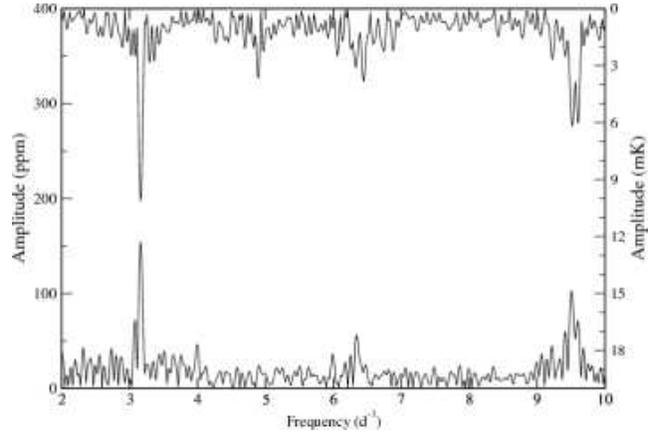}
			\caption{DFT spectra of the final $\gamma$\,Equ light curve (\em bottom\rm ) and the corresponding CCD temperature time series (\em top\rm ). The photometric data show a variation with a frequency of about $3.16$\,d$^{-1}$ (plus the first and second overtone) which is also visible in the DFT of the CCD temperature.}\label{f:CCDtemp}
		\end{figure}

		Before summer 2004, an on-board clocking problem caused a beat between the ACS and Science clocks. As a consequence of this beat, a spurious signal with a frequency of $\approx 3.16$\,d$^{-1}$ and harmonics of it are found in the MOST photometry obtained before mid 2004 (see Fig.\,\ref{f:CCDtemp}, bottom). This variation can also be found in a time series of the corresponding CCD temperature readings (Fig.\,\ref{f:CCDtemp}, top). The apparent temperature signal is an artefact of the timing beat and does not represent a real CCD temperature variation, so there is no correlation with CCD signal. Hence, this instrumental effect is eliminated by prewhitening frequency by frequency, being aware of the instrumental origin of the corresponding peaks in the amplitude spectrum.

		\subsubsection{Pointing stability}

\begin{figure}
			\includegraphics[height=240pt, angle=270]{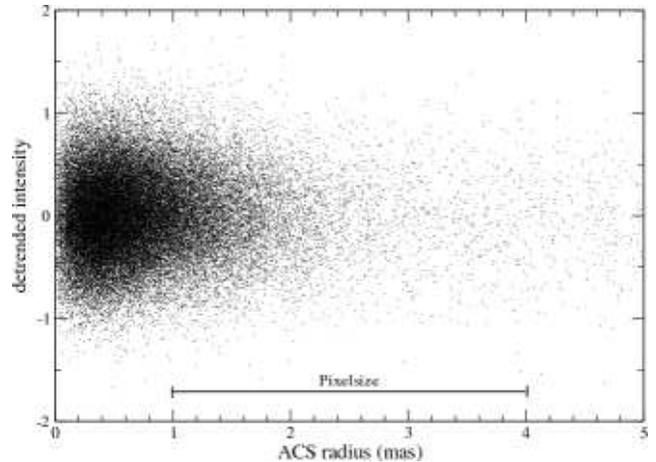}
			\caption{$\gamma$\,Equ photometry versus mean ACS error radius. The horizontal bar represents the physical pixel size of MOST.}\label{f:ACSmean}
		\end{figure}

		The ACS pointed the telescope at $\gamma$\,Equ with an average accuracy of about $\pm 1\farcs 8$, which is fairly below the physical pixel size of $3\arcsec$ and much better than what was achieved, e.\,g., for the first MOST science target, $\kappa ^1$\,Cet. As illustrated in Fig.\,\ref{f:ACSmean}, there is no correlation between the photometric data and the mean ACS error radius. The latter is the rms deviation of the line-of-sight from the ACS centre during one integration and an estimator for the pointing stability. Fig.\,\ref{f:ACSmean} also proves the homogeneity of the CCD sensitivity on a subpixel scale.

\section{Visualisation}					 			\label{sec:visual}
\begin{figure}
		\includegraphics[width=240pt]{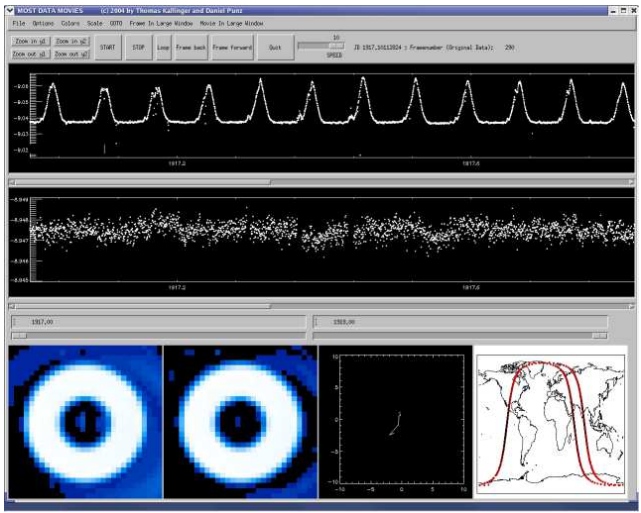}
		\caption{Screenshot of the MDM visualisation tool. The upper panels compare the raw and reduced light curves at different magnitude scales (in this case, $\eta$ Boo 2005). Scaling of time and magnitude, as well as animation range and speed are adjustable. Furthermore, the software provides monitoring of a variety of features in the lower panels (in this case, from left to right: raw SDS2 image, reduced SDS2 image, ACS error track, spacecraft track).}
		\label{fig:mdm}	
	\end{figure}

%xxxxxxxxxxxxxxxxxxxxx
	The IDL program {\sc Most Data Movies} (MDM) is a visualisation tool for SDS2 formatted data. In particular during the early attempts of understanding the instrumental properties it proved very helpful to ``see'' the effect of various reduction steps for individual frames as well as for pixel light curves. In addition, we wanted to correlate the MOST position with peculiarities in frames and light curves. For a convenient usage of MDM, several options are provided, like various scaling, color coding, speed adjustment for the animated frame sequence, and toggling stepwise back- or forward in time. The most helpful option for optimising the reduction routine is the simultaneous display of two data sets. For example, one can immediately compare the original with the corrected light curve. Fig.\,\ref{fig:mdm} shows the original and reduced light curves of \ge, the original and reduced Fabry Images (doughnuts) of a selected instance (marked by a pointer below the light curve) in the left bottom corner, next the ACS errors reported for this given exposure and the respective sub-satellite position on ground, colour coded (from red to black) with the magnetic field measured on-board.

	The MDM software is available on request.

\section{SDS1 data reduction}					 			\label{sec:SDS1}
%xxxxxxxxxxxxxxxxxx

The SDS1 data format represents row and column sums of pixel intensities rather than data resolved in two dimensions, which has several implications on the reduction procedure.
\begin{itemize}
\item There are only 8 binned pixels (i.\,e. the 2 pixels at each margin of row and column sums) that may safely be defined as background.
\item The image geometry may be treated as similar to SDS2 data (see \ref{ssec:shape}), but in general, local bumps in the doughnut cannot be resolved in the binned data.
\item Reliable automated identification of cosmic ray hits (according to \ref{ssec:cosmicscor}) is nearly impossble, because in the SDS1 pixel binning, one or a few pixels affected by a cosmic ray will only produce a small excess in the total integrated signal. Of course, this also means that most cosmic ray hits produce relatively small outliers in the MOST SDS1 photometry.
\item The restriction to 4 background pixels permits only 4 decorrelation steps in the stray-light correction procedure (as described in \ref{ssec:strayl}).
\end{itemize}
These aspects of SDS1 data limit the amount of processing possible on the ground, which was always recognised as the trade-off for being able to back up still useful photometry on board for many days to avoid possible gaps in the high-duty-cycle time series.

\begin{figure}
		\includegraphics[width=240pt]{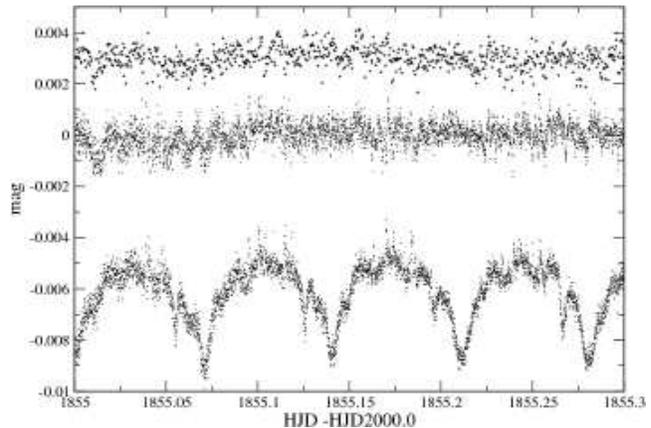}
		\caption{Comparison of the SDS1 and SDS2 light curves for Procyon 2005.
		{\em Bottom:} Raw SDS1 data. 
		{\em Mid:} Reduced SDS1 data.
		{\em Top:} Reduced SDS2 data.}
		\label{fig:sds1}	
	\end{figure}

Instead of applying the standard procedure described above to the SDS1 data, a more promising idea was to compute the differences between the reduced target intensity and the integrated intensity of the raw image for all SDS2 frames and to interpolate these differences in time to obtain an estimated background correction for the SDS1 measurements. The noise level of these is still expected to be higher than that of the SDS2 light curve, but the larger number of readings may nevertheless reduce the noise level of the Fourier amplitudes. As an example, in the Procyon 2005 photometry,
the number of data points in SDS2 format is only $48\, 840$, compared $280\, 995$ when SDS1 data are included. The corresponding amplitude noise level of SDS1\,$+$\,SDS2 data is reduced by about $20\,\%$ relative to SDS2 alone. Fig.\,\ref{fig:sds1} compares the SDS1 and SDS2 light curves for Procyon 2005. The fact that we gain only 20\,\% in terms of amplitude noise is due to the poorer quality of the $232\, 155$ SDS1 exposures.

\section{Less promising approaches}					 				\label{sec:dead}
%xxxxxxxxxxxxxxxxxx
While approaching what we think is now a well-developed reduction tool for space CCD photometry, we followed various ideas which we finally discarded, but which may be interesting in a different context, or may be traps for researchers in a similar situation as we were.

For example, we investigated various methods for stray-light correction. The simplest approach, similar to classical aperture photometry, was to subtract a mean background determined independently for each frame. Not surprisingly, this approach was too simple because of the complex stray-light pattern produced by the Fabry lens.

	\subsection{Local dispersion model}					\label{ssec:LDM}
	
\begin{figure}
		\includegraphics[width=240pt]{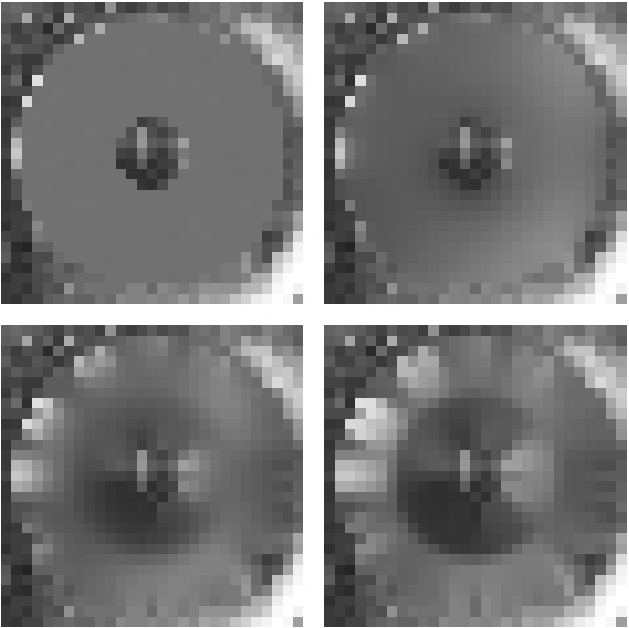}
		\caption{Background LDM models with different smoothing parameter $\lambda$ (0, 2, 5, and 10).}\label{f:ldm}
	\end{figure}

	The stray-light complication led us to develop a stray-light model taking the neighbouring background pixels into account. We used the spatial DWA $F_{\lambda\lambda}$ (according to eq.\,\ref{EQWhittakerSpace}) to describe the correlation between one pixel and its environment in the image and called this stray-light model a {\em Local Dispersion Model} (LDM). Fig.\,\ref{f:ldm} shows some LDM background models for the $\delta$\,Cet data with different smoothing parameters $\lambda$. Choosing $\lambda$ as the only free parameter did not lead to a satisfactory stray-light correction. 

	\subsection{Moment analysis}						\label{ssec:moment}
	
	Another stray-light correction scheme took advantage of an obvious direct correlation between the mean background and the target intensity, which suggested that one should compute the mean background as the normalised bivariate moment of order zero. A correction of the mean background by calculating a regression of the target intensity relative to a mean background and correcting for a slope led to clearly better results than the LDM described in \ref{ssec:LDM}. This encouraged us to investigate higher order correlations using
	\begin{equation}
	\langle ^{i}_{j}\rangle \equiv \frac{\langle l^jm^{i-j}I_{lm}\rangle _n}{\langle l^{2j}m^{2\left( i-j\right)}\rangle _n}\, ,
	\end{equation}
	where indices $l$, $m$ refer to background pixels only. The result is a bivariate polynomial fit $f$ of order $i$ to the background pixel intensities, which is obtained by setting
	\begin{equation}
	f_{lmn} \equiv \langle ^{i}_{j}\rangle \left( l-\langle l\rangle\right) ^j\left( m-\langle m\rangle\right) ^{i-j}\, .
	\end{equation}
	
\begin{figure}
		\includegraphics[width=240pt]{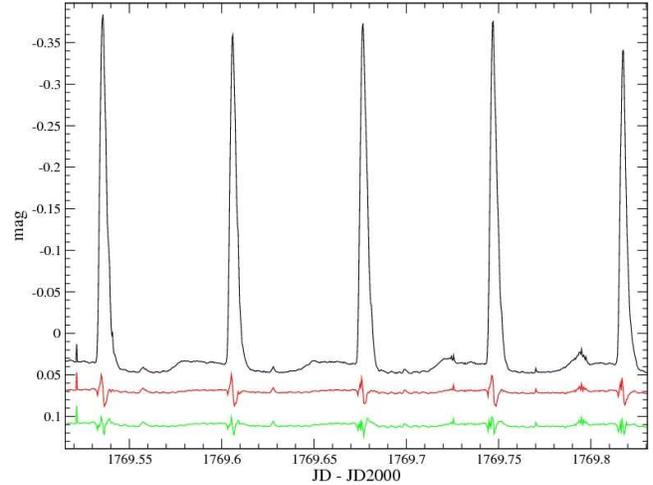}
		\caption{Part of the HR 1217 light curve without correction of background moments ({\em top}), after correction of zero order moment ({\em mid}), and after fifth order moment correction ({\em bottom}).}\label{f:moments}
	\end{figure}

	Applying various moment orders led to quite impressive results, but the stray-light peaks still dominated the light curve (Fig.\,\ref{f:moments}).
	
	\subsection{Fabry lens profile}				\label{ssec:fabryprof}

%	\begin{figure}
%		\caption{The path of a target on the Fabry lens surface during a single measurement may be recovered by the corresponding ACS $xy$-errors. Example for $\delta$\,Cet.}
%		\label{f:acspath}
%	\end{figure} 
	\begin{figure}
		\includegraphics[width=240pt]{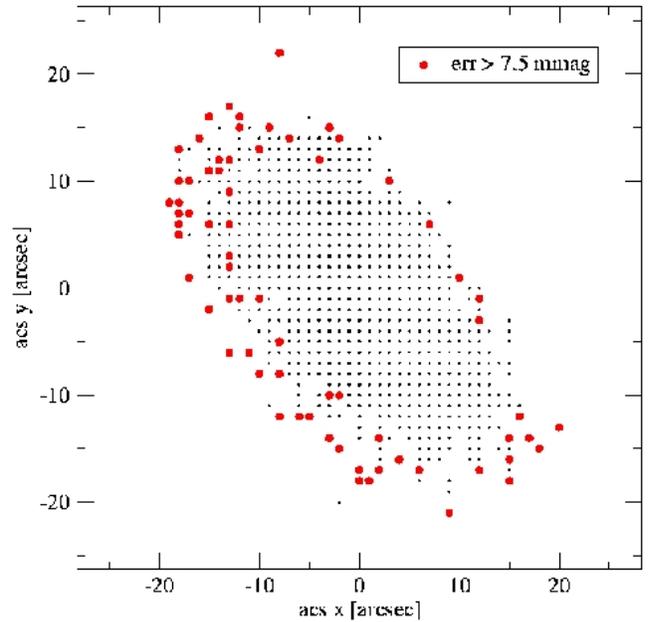}
		\caption{2D scan of Fabry lens 3/3. {\em Bullets:} ACS positions leading to corrupted data points.}
		\label{fig:flens}
	\end{figure} 
	
	The ``lens profile'' is intended to indicate optical distortions and disturbances by flawed areas on the Fabry lens. It is obtained by an analysis of individual MOST-images and allows estimating the influence of lens properties on the light curve as well as eventually finding an algorithm to correct for lens inhomogeneities.
	
	Impurities in the lens, like small inclusions or bubbles, will result in a reduced signal every time light passes through such blemishes. In a first step we simulated a constant light source, which was performed for $\delta$\,Cet by prewhitening the light curve with 20 frequencies. Then we correlated the residual intensity to the ACS error values stored in the image headers, indicating the deviation of the line-of-sight from the ideal position referring to the centre of the frame. 

	Usually there exists more than one ACS error value for an image, because the ACS system acquires a reading every second, while the typical integration time for a single frame is $10$ seconds and longer. In the case of $\delta$\,Cet ($10$ seconds integration time), up to $10$ ACS error values describe the movement of the target across the aperture during a given exposure. It is therefore impossible to immediately assign an erroneous data point to a specific ACS value. To overcome this problem, we used a grid of ACS error values and classified the boxes as ``good'' or ``bad'' according to a chosen threshold for the residual to the ``constant'' light curve. ``Good'' boxes are those where at least one data point has been obtained with a value below the chosen threshold and which contains an ACS value corresponding to the given box. ``Bad'' boxes obviously are those where no such frames are found. Changing the criterion from ``one'' good data point to ``$n$'' good data points for defining a good box does not change the result substantially.

	The lens profile based on data obtained for $\delta$\,Cet (Fig.\,\ref{fig:flens}) shows that the given Fabry lens is perfect except for the outermost area, where the ACS is at its limit. The pointing precision has significantly improved since the beginning of MOST observations. Now it is practically impossible to produce a reliable lens profile using this technique. However, in case the ACS gets worse and lenses get damaged we might have to return to this tool for properly identifying bad data points. For the time of this report and as a conclusion, we do not have indications of Fabry lenses influencing the data quality.

\section{Results}
	
For $13$ MOST targets (Table \ref{TABquality}) observed from October, 2003, to March, 2005, the typical fraction of exposures transmitted in the SDS2 format was 41\,\%. About 52\,\% of the pixels in a Fabry frame were initially used for defining the target aperture mask. Out of all SDS2 exposures, $3.7\pm 1.8$\,\% had to be rejected due to ACS problems (\ref{ssec:rejection}), and $0.8\pm 0.8$\,\% due to a distorted Fabry Image geometry (\ref{ssec:shape}). Close to 0.47\,\% of all SDS2 exposures contained pixels apparently affected by cosmic rays (\ref{sssec:coscand}), but only 47\,\% of these conspicuous pixels finally turned out to be corrupted by cosmic rays (\ref{sssec:cosaffect}). 5\,\% of SDS2 exposures had to be rejected due to an excessive cumulation of cosmic rays (\ref{sssec:cosrates}). Between 18 and 224 decorrelation steps were used for the final reduction (\ref{sssec:decorrelation}) with typically 56\,\% of the pixels defined in the initial target aperture mask used for the resulting light curves (\ref{ssec:pixsel}).
		
\section{Conclusions}					 				\label{sec:concl}

The reduction method described in this paper was applied to MOST Fabry Image photometry and relies on the following steps:
\begin{itemize}
	\item construction of a ``data cube'' of all CCD frames including all FITS header information to which all software components consistently refer
	\item definition of target and background pixels
	\item rejection of images deviating significantly from the average
	\item cosmic ray correction (correction of individual pixels or elimination of an entire image, if the number of pixels to be corrected is above a chosen threshold)
	\item stray-light correction - the core of our reduction tool. This correction is based on the assumption that the stray-light sources superpose for a given pixel as individual point sources. This property allows one to compensate for the orbit-modulated stray light by decorrelation of background and target pixel intensities -- thus avoiding any period folding techniques. The amplitudes of orbit harmonics are reduced by up to four orders of magnitude.    
	\item compensation for reduced intrinsic amplitudes due to repeated decorrelations
	\item selection of best suited pixel light curves for merging into the final target star light curve
	\item check for remaining correlation of photometry with spacecraft parameters
\end{itemize}
The software package is very efficient and provides the reduction of a large amount of data in a pipe-line style. The improvement in data quality achievable with our method is displayed in Table \ref{TABquality} (p.\,\pageref{TABquality}) containing a comparison of both stray-light amplitude and frequency-domain noise for raw vs. reduced data. A comparison with other reduction techniques clearly indicates that no artefacts are introduced in the photometry. This is a considerable advantage when interpreting complex frequency spectra with amplitudes close to the noise level.

Simple binning of consecutive orbits and a corresponding trendline correction will perform at least as well as the decorrelation technique in the sense of noise. But the noise level is not the only (and also not the best) quality estimator. Time-domain binning leads to the introduction of a periodic frequency-domain filter function. In other words, the spectral noise level is not uniformly distributed in frequency, which has serious implications on the detailed analysis of individual frequencies. The noise level can be pushed down considerably while paying the price of distortion, which we consider less convenient than the slightly higher, but undistorted noise. Although there are frequency regimes and timescales of intrinsic stellar variation where the better photon statistics of binned data may be more desirable for the task at hand than e.\,g. eliminating pixels from the Fabry image and increasing the overall Poisson noise, we are generally convinced that a low noise level is not the only thing to strive for. This is why we completely omit manipulations of the total light curve (like binning of orbits) without carefully examining the origin of the contamination we correct for.
	
\section*{Acknowledgments}
D.\,F., M.\,G., D.\,H., T.\,K., D.\,P., P.\,R., S.\,S., and W.\,W.\,W. received financial support from the Austrian FFG -- Areonautics and Space Agency -- and by the Austrian Science Fonds (FWF-P17580). The Natural Sciences and Engineering Research Council of Canada supports the research of D.\,B.\,G., J.\,M.\,M., A.\,F.\,J.\,M., S.\,M.\,R., and G.\,A.\,H.\,W.; A.\,F.\,J.\,M. is also supported by FCAR (Quebec). R.\,K. is supported by the Canadian Space Agency.
The MOST ground station in Vienna was developed and is operated in cooperation with the Vienna University of Technology (W. Keim, A. Scholtz, V. Kudielka).

%\label{lastpage}

\end{document}